\documentstyle[aps,prb,eqsecnum,twocolumn,epsf]{revtex}
\begin{document}
\draft
\title{Mean parameter model for the Pekar-Fr\"{o}hlich polaron \\ in
a multilayered heterostructure}
\author{M.\ A.\ Smondyrev$^{a)}$\cite{x1}, B.\ Gerlach$^{b)}$,
M.\ O.\ Dzero$^{a)}$\cite{x2}}
\address{$^a)$ Bogoliubov Laboratory of Theoretical Physics, Joint Institute
for Nuclear Research\\ 141980 Dubna, Moscow Region, Russia}
\address{$^b)$ Institut f\"ur Physik, Universit\"at Dortmund,
44221 Dortmund, Germany}
\date{\today}
\maketitle
\begin{abstract}
The polaron energy and the effective mass are calculated for an electron
confined in a finite quantum well constructed of
$GaAs/Al_x Ga_{1-x} As$ layers. To simplify the study we suggest
a model in which parameters of a medium are averaged over
the ground-state wave function. The rectangular and the Rosen-Morse
potential are used as examples.

To describe the confined electron properties
explicitly to the second order of perturbations in powers of
the electron-phonon coupling constant
we use the exact energy-dependent Green function
for the Rosen-Morse confining potential.
In the case of the rectangular potential, the sum over all
intermediate virtual states is calculated.
The comparison is made with the often used leading term approximation
when only the ground-state is taken into account as a virtual state.
It is shown that the results are quite
different, so the incorporation of all virtual states and especially
those of the continuous spectrum is essential.

Our model reproduces the correct three-dimensional asymptotics
at both small and large widths.
We obtained a rather monotonous behavior
of the polaron energy as a function of the confining potential width
and found a peak of the effective mass.
The comparison is made with theoretical results by other authors.
We found that our model gives practically
the same (or very close) results
as the explicit calculations for potential widths
$L \geq 10 \AA$.
\end{abstract}
\narrowtext

\section{Introduction}
Quasi-two-dimensional (2D) systems have attracted a lot of attention
during the last decade because of their practical realization. If
a heterostructure is made of polar materials such as $GaAs$,
$InAs$ etc., the electron-phonon interaction
modifies the properties of the electron confined
to a 2D-structure resulting in a shift of the binding energy
and the effective band mass.

The polaron effects in a 2D electron gas have
extensively been studied. At earlier stages the attention was paid
to the properties of a polaron confined to an infinite thin
2D layer\cite{sarma,pwd,selugin}. The binding energy and the effective
mass were calculated for a $GaAs/Al_xGa_{1-x}As$  infinitely deep
quantum well of a finite length\cite{sarma2,sarma3}.
In these papers only the interaction with the bulk LO-phonon
mode has been taken
into account. Actually, LO-phonon modes are modified in a 2D layer
(the so-called confined slab LO-phonon modes). Besides,
there exist interface optical-phonon modes as well as half-space
LO-phonon modes in a barrier
material\cite{firsov,evrard,fomin,wendler1,wendler2}.
For the review of these modes (also in complicated multi-layer
structures) see the book by Pokatilov, Fomin and Beril\cite{pfb}
and also more recent publications \cite{fomin2,fomin3} of this group.
The influence of the mentioned modes on polarons were studied in
Refs.~\onlinecite{licari,liang,cosmas,degani}.

While different phonon modes were studied in details,
the quantum well potential was supposed to be infinitely deep
in the cited papers. On the other hand, the properties of the
system would be quite different if a confining potential
had a finite depth. Indeed, for an infinitely deep confining potential
the binding energy is the monotonous function of the potential width
which varies between limiting values
$E^{(in)}_{3D} = \alpha_{in}\hbar\omega_{in}$
for the three-dimensional (3D) space and
$E^{(in)}_{2D} = (\pi/2)\,\alpha_{in}\hbar\omega_{in}$,
where $\alpha_{in}$ is the standard Fr\"ohlich electron-phonon coupling
constant and $\omega_{in}$ is the LO-phonons frequency
for the quantum well material.
If a particle is confined to a finite potential well, the limiting
value of the binding energy should be the same at large width
of the well. But when the width becomes too small, the energy level
approaches the edge of the well, so that effectively the particle
is spread over the 3D space. Thus, the limiting value of the binding
energy should coincide with the 3D limiting value rather than with
the 2D one. This means, the binding energy takes the value
$E^{(out)}_{3D} = \alpha_{out}\hbar\omega_{out}$ at small
widths where the parameters $\omega_{out}$ and
$\alpha_{out}$ are now related to the barrier material.
The binding energy evidently has a peak
at some intermediate value of the width if $E^{(out)}_{3D} \leq
E^{(in)}_{3D}$. If this is not the case, the existence of the peak
should be checked in more detail.

Different rectangular quantum wells of a finite height have been
investigated by Hai, Peeters and Devreese\cite{hpd2,hpd3}
and Shi, Zhu et al.\cite{china}
in the scope of the second order perturbation theory in
powers of the electron-phonon coupling constant $\alpha$ with
all phonon modes being incorporated.
Peaks of the phonon induced energy shift
and the polaron effective mass were found for some values
of the confining potential widths.

In principle, the same approach can be used while dealing with
a quantum well constructed of layers of different materials.
But the problem becomes then too complicated because one
has to take into account interface phonon modes at each
frontier of different materials as well as quantized phonon
modes inside each of the layers. The main goal of the present paper is to
formulate a simplified model to take these effects into account
and to deal with the effective confining potential and only one
bulk phonon mode. We calculate polaron characteristics for
the same rectangular quantum well as in Refs.~\onlinecite{hpd2,hpd3,china}
to compare the results. Another example is given
of a quantum well of the finite depth for which the second-order
correction due to the electron-phonon interaction can
be calculated explicitly. Namely, we take the Rosen-Morse
potential to confine electrons to a 2D-multilayered heterostructure
and calculate the shift of the ground-state energy and the effective mass
perturbatively, that is, in the weak-coupling limit.
In contrast with the rectangular potential we should not worry
about the correct including of all virtual states because the Green
function is known analytically for this system.

\section{Formulation of the model}
Let us consider a quantum well in the $z$ direction constructed
of the $xy$ plane layers of $GaAs/Al_xGa_{1-x}As$. That is,
the $AlAs$ mole fraction $x$ depends on the coordinate $z$: $x=x(z)$.
The energy gap between different materials forms the
confining potential $V(z)$ which serves us as the main entity.
Given the potential $V(z)$, one can find the corresponding
mole fractions $x(z)$ and a dependence on $z$ of any of
the medium parameters
(such as the electron band mass $m(z)$, phonon frequencies $\omega (z)$,
dielectric constants $\varepsilon_0 (z), \varepsilon_{\infty}(z)$,
Fr\"ohlich coupling constants $\alpha (z)$, etc.).

To avoid difficulties with mass mismatch in different layers
we suggest to use a {\em mean} band mass $m$ which is common for all layers.
Then we start with the following expression for the electronic part
of the Hamiltonian:
\begin{eqnarray}
H_{el} &=& H_{el,\parallel} + H_{el,\perp}, \nonumber \\
H_{el,\parallel} &=& {\vec p_{\parallel}^{\,2}\over 2m}, \quad
H_{el,\perp} =
{p_z^{\,2}\over 2m} + V(z),
\label{eq2.01}\end{eqnarray}
where the electron mean band mass is defined by the relation
\begin{eqnarray}
{1\over m} = \int dz\,{|\psi_1(z)|^2\over m(z)}
\label{eq2.02}\end{eqnarray}
and the ground state wave function $\psi_1(z)$
for the electron motion in $z$ direction
is a solution to the Schr\"{o}dinger equation
\begin{eqnarray}
H_{el,\perp}\psi_1 = E_1 \psi_1
\label{eq2.03}\end{eqnarray}
with $E_1$ being a ground state energy.
As the wave function $\psi_1$ also depends on the mean
band mass $m$, the latter can be found as
a self-consistent solution of Eqs.~(\ref{eq2.02}), (\ref{eq2.03}).

In a similar way we define the free LO-phonon Hamiltonian
\begin{eqnarray}
H_{ph} = \hbar \omega_{\mbox{\tiny LO}}
\sum\limits_{\vec k} a^{\dag}_{\vec k}a_{\vec k},
\label{eq2.04}\end{eqnarray}
where $a^{\dag}_{\vec k}\;(a_{\vec k})$ are creation (annihilation)
operators of a phonon with a wave vector $\vec k$,
and mean frequency $\omega_{\mbox{\tiny LO}}$ can be found
from the expression
\begin{eqnarray}
\omega_{\mbox{\tiny LO}}= \int dz\,\omega(z)\,|\psi_1(z)|^2.
\label{eq2.05}\end{eqnarray}
Evidently, we have to address why the free phonon Hamiltonian is
averaged with respect to the electron wave function. Our motivation
is based on the fact that we are going to apply our model to calculate
polaron effects. That is, our {\em effective} phonons will be considered
only in a cloud around the electron, and the properties of this cloud
depend on the electron position. So, in our model
the effective phonons replace
numerous phonon modes whose frequencies depend on the coordinate
$z$ of the electron.

Finally, we accept the conventional form
of the Hamiltonian describing the interaction
of the electron with effective phonons:
\begin{eqnarray}
H_{int} = \sum_{\vec k}\,\left(a_{\vec k}\,V_{\vec k}\,
e^{i \vec k \cdot \vec r}   +
a^{\dag}_{\vec k}\,V^{*}_{\vec k}\,e^{-i \vec k \cdot \vec r}\right),
\label{eq2.06}\end{eqnarray}
where the Fourier transforms of the electron-phonon interaction
potential are specified as follows:
\begin{eqnarray}
V_{\vec k} = -i\hbar \omega_{\mbox{\tiny LO}} \left( {4\pi\alpha \over Vk^2}
\sqrt{{\hbar \over 2m\omega_{\mbox{\tiny LO}}  }} \right)^{1/2}.
\label{eq2.07}\end{eqnarray}
Here the use is made of a mean Fr\"{o}hlich coupling constant $\alpha$
which can be found from the relation
\begin{eqnarray}
\sqrt{\alpha}
= \int dz\, |\psi_1(z)|^2\,{\omega(z)\over
\omega_{\mbox{\tiny LO}}} \left( \alpha(z)
\sqrt{m\omega_{\mbox{\tiny LO}}\over m(z)\omega(z) } \right)^{1/2}.
\label{eq2.08}\end{eqnarray}
Note that we define the mean parameters in
Eqs. (\ref{eq2.02}), (\ref{eq2.05}), (\ref{eq2.08})
according to the way they enter the Hamiltonian.

Thus, we describe a complicated multilayered heterostructure
by the Hamiltonian
\begin{eqnarray}
H = H_{el} + H_{ph} + H_{int}
\label{eq2.09}\end{eqnarray}
with the bulk phonon mode only which inhabits an effective medium
with mean characteristics defined above.
The details of the heterostructure are taken
into account by the confining potential $V(z)$.

Performing a unitary transformation $H \to H'= U^{-1}HU$
with the operator
\begin{eqnarray}
U= \exp\left[ -i\vec r_{\parallel}\sum_{\vec k} \vec k_{\parallel}
a^{\dag}_{\vec k}a_{\vec k}\right],
\label{eq2.10}\end{eqnarray}
we arrive at the Hamiltonian
\begin{mathletters}
\begin{eqnarray}
H' =  H'_{el,\parallel} + H_{el,\perp} + H_{ph} + H'_{int},
\label{eq2.11a}
\end{eqnarray}
\begin{eqnarray}
H'_{el,\parallel} &=& {1\over 2m}\left(\vec p_{\parallel} - \hbar \sum_{\vec k}
\vec k_{\parallel} a^{\dag}_{\vec k}a_{\vec k}\right)^2,
\label{eq2.11b}
\end{eqnarray}
\begin{eqnarray}
H'_{int} = \sum_{\vec k}\,\left(a_{\vec k}\,V_{\vec k}\,
e^{i k_z \cdot z}   +
a^{\dag}_{\vec k}\,V^{*}_{\vec k}\,e^{-i k_z \cdot z}\right).
\label{eq2.11-4}\end{eqnarray}
\label{eq2.11}\end{mathletters}
The quantity $\vec p_{\parallel}$ is a c-number corresponding to the
conserved momentum in the $xy$ plane and the Hamiltonians
$H_{el,\perp}, H_{ph}$ are defined by Eqs.~(\ref{eq2.01}), (\ref{eq2.04}),
respectively.

Keeping in mind the smallness of the electron-phonon coupling constant
$\alpha$ for most of the materials, we calculate the second-order
correction to the unperturbed Hamiltonian $H'_0 =
H'_{el,\parallel} + H_{el,\perp} + H_{ph}$
(note that the quantum-mechanical first-order correction is equal to zero).
The unperturbed energy levels are given by the expression
\begin{eqnarray}
E(\vec p_{\parallel},n_{\vec k},N) &=&
{1\over 2m}\left(\vec p_{\parallel} - \hbar\sum_{\vec k}\vec k_{\parallel} n_{\vec k}\right)^2
+  \nonumber \\
&& \hbar \omega_{\mbox{\tiny LO}} \sum_{\vec k}\, n_{\vec k} +E_N,
\label{eq2.12}\end{eqnarray}
where $n_{\vec k}$ is the number of phonons with the wave vector $\vec k$.
The energy $E_N$ is the $N$-th energy level of the one-dimensional
system $H_{el,\perp}$ of Eq.~(\ref{eq2.01}).
Here $N$ is the corresponding quantum number
not necessarily a discrete one:
it stands for both the quantum number $n$ which varies from 1 to $n_{max}$
and the wave vector $q$ of the continuous spectrum states.

The wave functions of the unperturbed Hamiltonian $H'_0$ are given
by the direct product
\begin{eqnarray}
|\vec p_{\parallel};n_{\vec k},N \rangle =
|n_{\vec k}\rangle \otimes |N\rangle
\label{eq2.13}\end{eqnarray}
of the corresponding wave functions of different terms in $H'_0$.

Because of the structure of the interaction term $H'_{int}$
only intermediate states with one phonon contribute to the second
order correction to the ground-state energy. The latter is then given
by the expression
\begin{eqnarray}
&& \Delta_2 E(\vec p_{\parallel}) = \nonumber \\
&&- \sum\limits_{N,\vec k}{|V_{\vec k}|^2\ |G(N,k_z)|^2 \over
E_N + \hbar \omega_{\mbox{\tiny LO}} + \displaystyle
{(\vec p_{\parallel} - \hbar \vec k_{\parallel})^2
- {\vec p_{\parallel}}^{\,2} \over 2m}-E_1},
\label{eq2.14}\end{eqnarray}
where
\begin{eqnarray}
G(N,k_z) = \int\limits_{-\infty}^{\infty} dz\,\psi_{N}(z)\psi_{1}(z)
e^{i k_z z}
\label{eq2.15}\end{eqnarray}
and $\psi_N(z)$ are the wave functions of the Hamiltonian $H_{el,\perp}$
in Eq.~(\ref{eq2.01}).
The concrete application of these formulae is given in the
following section.

\section{Rectangular Potential}
\subsection{Medium mean characteristics}
As an example we now consider the rectangular confining potential
\begin{eqnarray}
V(z) = \left\{
\begin{array}{ll}
0,\, & |z| \leq L/2  \\
V_0, & |z| > L/2
\end{array}  \right.
\label{eq3.01}\end{eqnarray}
and
\begin{eqnarray}
m(z) = \left\{
\begin{array}{ll}
m_{in},\, & |z| \leq L/2  \\
m_{out}, & |z| > L/2
\end{array}  \right.
\label{eq3.02}\end{eqnarray}
with $m_{in}\ (m_{out})$ being the electron band masses
in the well (barrier) material, respectively.
For concreteness we assume $GaAs$ to be the quantum well material
and $Al_xGa_{1-x}As$ to be the barrier material.

Symmetrical wave functions of the discrete spectrum in the rectangular
quantum well with the mean band mass $m$ take the form
\begin{eqnarray}
\psi_{s,n} = N_{s,n}\left\{
\begin{array}{ll}
\cos q_nz , & |z| \leq L/2 \\
\cos \displaystyle {q_n L\over 2}
e^{- p_n (|z|-L/2)}, & |z|>L/2,
\end{array} \right.
\label{eq3.03}\end{eqnarray}
where
\begin{eqnarray}
p_n = \sqrt{q_{max}^2 - q_n^2}, \quad q_{max}^2 = {2mV_0\over \hbar^2}
\label{eq3.04}\end{eqnarray}
and the normalization constant is given by
\begin{eqnarray}
N_{s,n} = \sqrt{ 2 p_n\over p_n L + 2}.
\label{eq3.05}\end{eqnarray}

Antisymmetrical wave functions of the discrete spectrum take the form
\begin{eqnarray}
&&\psi_{a,n} = N_{s,n}\left\{
\begin{array}{ll}
\sin q_nz , & |z| \leq L/2 \\
\mbox{\rm sgn}(z)\, \sin \displaystyle {q_n L\over 2}
e^{- p_n (|z|-L/2)}, & |z|>L/2
\end{array} \right. \nonumber \\
&&
\label{eq3.06}\end{eqnarray}
with the same normalization constant given by Eq. (\ref{eq3.05}).

The total number $n_{max}$ of the discrete energy levels is given
by the expression
\begin{eqnarray}
n_{max} = 1 + \left[{q_{max} L\over \pi}\right],
\label{eq3.07}\end{eqnarray}
where $[A]$ is an integer part of a number $A$.
The expression for the discrete energy levels reads as follows:
\begin{eqnarray}
{q_nL\over 2} = \mbox{\rm atan}\,
\sqrt{{q_{max}^2\over q_n^2} -1} +{\pi (n-1)\over 2},\quad n=1,2,\ldots .
\label{eq3.08}\end{eqnarray}
Energies with odd (even) $n$ correspond to the symmetrical
(antisymmetrical) wave functions.

The energy $E_q = \hbar^2 q^2/2m$ of the continuous spectrum state
depends on the wave vector $q$. The corresponding symmetrical
wave functions are as follows:
\begin{eqnarray}
&& \psi_{s,q} = {N_{s,q}\over \sqrt{L_z}} \left\{
\begin{array}{ll}
 p\cos qz,  & |z| \leq L/2, \\[5mm]
  p\cos \displaystyle {qL\over 2}
\cos \displaystyle p(|z|-L/2) - \\[2mm]
 q \sin \displaystyle {qL\over 2} \sin p(|z|-L/2), & |z| > L/2, \\
\end{array} \right. \nonumber \\
&&
\label{eq3.09}\end{eqnarray}
where
\begin{eqnarray}
p=\sqrt{q^2 - q^2_{max}}
\label{eq3.10}\end{eqnarray}
and $L_z$ is the (infinite) size of the system in the $z$ direction.
The normalization constant is given by the expression
\begin{eqnarray}
N_{s,q}= \sqrt{2\over p^2 \cos^2 \displaystyle {qL\over 2}
+  q^2 \sin^2 \displaystyle {qL\over 2}}.
\label{eq3.11}\end{eqnarray}
The antisymmetrical wave functions are as follows:
\begin{eqnarray}
&& \psi_{a,q} = {N_{a,q}\ \mbox{\rm sgn}(z) \over \sqrt{L_z}} \left\{
\begin{array}{ll}
p\sin q|z|, \quad |z| \leq L/2, \\[5mm]
 p\sin \displaystyle {qL\over 2}
\cos p(|z|-L/2) + \\[2mm]
 q \cos \displaystyle {qL\over 2} \sin p(|z|-L/2), \\ \qquad\qquad\qquad
|z| > L/2,
\end{array} \right. \nonumber \\
&&
\label{eq3.12}\end{eqnarray}
where the normalization constant is given by the expression
\begin{eqnarray}
N_{a,q}= \sqrt{2\over p^2 \sin^2 \displaystyle {qL\over 2}
+ q^2 \cos^2 \displaystyle {qL\over 2}}.
\label{eq3.13}\end{eqnarray}

The electron mean band mass is defined as
\begin{eqnarray}
{1\over m} &=& {W_{in}\over m_{in}} + {W_{out}\over m_{out}}
\rightarrow \nonumber \\
m &=& {m_{in}m_{out}\over W_{in} m_{out} +(1-W_{in})m_{in}},
\label{eq3.14}\end{eqnarray}
where $W_{in}$ and $W_{out}=1-W_{in}$ are probabilities to find the electron
inside (outside) the quantum well. The expression for $W_{in}$ follows
from Eq.~(\ref{eq3.03})
\begin{eqnarray}
W_{in} = 2N^2_{s,1}\int\limits_0^{L/2}dz\,\cos^2 q_1 z =
1- {(q_1/q_{max})^2 \over 1+ p_1 L/2},
\label{eq3.15}\end{eqnarray}
where $q_1$ is a solution to Eq.~(\ref{eq3.08}) for the ground-state ($n=1$).

To finish this subsection, we note that the exact
energy levels in the rectangular potential with different
masses $m_{in}$ and $m_{out}$ calculated for the
$GaAs/Al_x Ga_{1-x} As$ heterostructure
practically coincide with the levels
obtained with the electron mean band mass $m$.
To obtain an inner criterion of the validity of the anzatz
concerning the mean band mass we notice that the particle being
on lowest energy levels is located mostly inside the well which
means that its band mass is almost coincide with $m_{in}$.


\noindent
One can
await the largest discrepancy for a level near the potential edge.
The $n$-th discrete level appears at the width $L=L^{(av)}_n$,
where
\begin{eqnarray}
L^{(av)}_n = \pi(n-1){\hbar \over \sqrt{2mV_0}} = {\pi (n-1)\over q_{\max}},
\label{eq3.16}\end{eqnarray}
and the analogous width for the exact solution reads as follows:
\begin{eqnarray}
L^{(ex)}_n = \pi(n-1){\hbar \over \sqrt{2m_{in}V_0}}.
\label{eq3.17}\end{eqnarray}


Thus, the ratio
\begin{eqnarray}
{L^{(av)}_n\over L^{(ex)}_n} = \sqrt{m_{in}\over m} =
\sqrt{W_{in} + (1-W_{in}){m_{in}\over m_{out}}}.
\label{eq3.18}\end{eqnarray}
can serve us as the numerical criterion of the validity
of the anzatz. The largest discrepancy happens at $n=2$ and
in this case Eqs. (\ref{eq3.04}), (\ref{eq3.08}) ¨ (\ref{eq3.15})
lead to the following expression:
\begin{eqnarray}
{L^{(av)}\over L^{(ex)}} = \sqrt{0.844+ 0.156 {m_{in}\over m_{out}}}.
\label{eq3.19}\end{eqnarray}
Note that numerical coefficients here do not depend on
the material parameters. For the $GaAs/Al_{0.3} Ga_{0.7} As$
quantum well we have $m_{in}/m_{out} \approx 0.7$ and
the discrepancy is about $2\%$; in the worst possible case when
$m_{in}/m_{out} \ll 1$ the discrepancy is still not large:
$100\% \sqrt{0.844} \approx 8\%$.

\subsection{Electron-phonon correction to the polaron energy and the
effective mass}

Summation over the wave vector $\vec k$ in Eq.~(\ref{eq2.14})
can be reduced to integration in a conventional way
\begin{eqnarray}
&& \sum\limits_{\vec k} |V_{\vec k}|^2\ (\ldots) =
{V\over (2\pi)^3}\int d{\vec k}\,|V_{\vec k}|^2\ (\dots) = \nonumber \\
&& (\hbar\omega_{\mbox{\tiny LO}})^2 \sqrt{\hbar \over 2m \omega_{\mbox{\tiny LO}}}
\,{\alpha\over 2\pi^2}
\int {d\vec k_{\parallel}\, dk_z \over k^2_{\parallel} + k^2_{z}}\
(\ldots ).
\label{eq3.20}\end{eqnarray}
Then, the integration over $\vec k_{\parallel}$ in Eq.~(\ref{eq2.14})
can be performed explicitly. As we are interested in corrections
to the ground-state energy and the effective mass
$m_{eff} \approx m + \Delta_2 m$ of the polaron motion in the $xy$ plane,
we expand $\Delta_2 E(\vec p_{\parallel}) \approx
\Delta_2 E - \displaystyle
{\Delta_2 m \over 2m^2}\,{\vec p}\,^2_{\parallel}$
in powers of the conserved momentum $\vec p_{\parallel}$.
Doing this the use is made of the integral
\begin{eqnarray}
&& \int {d^2 \vec k_{\parallel} \over ({\vec k}^2_{\parallel} + k^2_{z})
[{\vec k_{\parallel}}^{\,2} -
2 \vec k_{\parallel}\cdot \vec p_{\parallel}/\hbar + b^2]} \approx
\pi {\ln (k^2_{z}/b^2) \over k^2_{z} - b^2} + \nonumber \\
&& \left({\vec p_{\parallel} \over \hbar}\right)^2\,\pi\,
{k^4_{z} -b^4 -2 k^2_{z} b^2 \ln (k^2_{z}/b^2)
\over b^2 (k^2_{z}-b^2)^3}.
\label{eq3.21}\end{eqnarray}

As the next step we use dimensionless ``polaronic" units
performing the scaling
$ k_z \to k_z\sqrt{2m\omega_{\mbox{\tiny LO}}/\hbar}, \
z \to z \sqrt{\hbar / 2 m\omega_{\mbox{\tiny LO}}}$
and using the notation
\begin{eqnarray}
l=L \sqrt{2m\omega_{\mbox{\tiny LO}}\over \hbar}, \quad
\varepsilon_N = {E_N\over \hbar\omega_{\mbox{\tiny LO}}}.
\label{eq3.22}\end{eqnarray}
In these units the correction to the ground-state energy takes the form
\begin{eqnarray}
&&{\Delta_2 E\over \hbar \omega_{\mbox{\tiny LO}}} =   \nonumber \\
&& - {\alpha\over \pi}
\,\sum\limits_{N} \int\limits_0^{\infty} dk_z
{\ln (k^2_{z}/b^2_N) \over k^2_{z} - b^2_N}
\left(|G_s (N,k_z)|^2 + |G_a (N,k_z)|^2\right),  \nonumber \\
&&
\label{eq3.23}\end{eqnarray}
where
\begin{eqnarray}
b_N = \sqrt{\varepsilon_N +1-\varepsilon_1}.
\label{eq3.24}\end{eqnarray}
The correction to the effective mass reads as follows:
\begin{eqnarray}
&&{\Delta_2 m\over m} = \nonumber \\[3mm]
&&{\alpha\over \pi}
\sum\limits_{N} \int\limits_0^{\infty} dk_z
{k^4_{z} -b^4_N -2 k^2_{z} b^2_N \ln (k^2_{z}/b^2_N)
\over b^2_N (k^2_{z}-b^2_N)^3} \times  \nonumber \\
&& \left(|G_s (N,k_z)|^2 + |G_a (N,k_z)|^2\right).
\label{eq3.25}\end{eqnarray}
Quantities $G_j(N,k_z)$ in Eqs.~(\ref{eq3.23}), (\ref{eq3.25}) are given in
dimensionless units by the same Eq.~(\ref{eq2.15}); the indices $(a)s$
are related to  (anti)symmetrical wave functions being used in Eq. (\ref{eq2.15}):
\begin{eqnarray}
G_s(N,k_z) = 2\int\limits_0^{\infty} dz\,\psi_{s,N}(z)\psi_{s,1}(z)\,\cos k_zz,
\nonumber \\
G_a(N,k_z) = 2\int\limits_0^{\infty} dz\,\psi_{a,N}(z)\psi_{s,1}(z)\,\sin k_zz,
\label{eq3.26}\end{eqnarray}

Evidently, the replacement $L \to l$
should be done in the definition of the wave functions
and their normalization constants; in addition $L_z \to l_z$ in
Eqs.~(\ref{eq3.09}), (\ref{eq3.12}) as well as
in Eq.~(\ref{eq3.08})
for the energy levels of the discrete spectrum.
Eq.~(\ref{eq3.04}) now reads as follows:
\begin{eqnarray}
p_n = \sqrt{v_0 - q_n^2}, \quad v_0 = {V_0\over \hbar\omega_{\mbox{\tiny LO}}},
\quad q^2_{max} = v_0.
\label{eq3.27}\end{eqnarray}
Eq.~(\ref{eq3.07}) takes the form
\begin{eqnarray}
n_{max} = 1+\left[{\sqrt{v_0}l \over \pi}\right].
\label{eq3.28}\end{eqnarray}
The relation of dimensionless energies of the discrete and continuous
spectra with subsequent wave vectors takes the form
$\varepsilon_n = q^2_n,\ \varepsilon_q = q^2$.
All the changes mentioned should also be done in Eq.~(\ref{eq3.15}).

The final note of this section concerns summation over $N$ in
Eqs.~(\ref{eq3.23}), (\ref{eq3.25}):
\begin{eqnarray}
\sum\limits_{N}\,(\ldots) = \sum_{n=1}^{n_{max}}\,(\ldots) +
\lim_{l_z\to \infty}
{l_z\over 2\pi} \int\limits_0^{\infty}\!\! dp\,(\ldots).
\label{eq3.29}\end{eqnarray}
The replacing of the sum over the continuous spectrum
by the integration over the wave vector $p$ follows from
Eqs. (\ref{eq3.09}, \ref{eq3.12}) in the limit $L_z \to \infty$.
The wave vectors $q$ and $p$ are related to each other because of
Eq.~(\ref{eq3.10})
which now takes the form $q=\sqrt{p^2+v_0}$. Note also that
only $G_s(N, k_z)$ ($G_a(N, k_z)$) has to be taken into account
for odd (even) $n$ in the sum over the discrete quantum number $n$.

The numerical results obtained are plotted in Fig.~\ref{fig1}
for $\Delta_2 E$ and in Fig.~\ref{fig2}a for $\Delta_2 m/m$.
Because the mean effective mass $m$ depends on the potential width
we also plotted in Fig.~\ref{fig2}b
the ratio of the mass shift to
those in the well material, that is, the ratio
\begin{eqnarray}
\delta_2 m = {\Delta_2 m\over \Delta_2 m_{in}}, \quad
\Delta_2 m_{in} = m_{in}{\alpha_{in}\over 6}.
\label{eq3.30}\end{eqnarray}
The discussion of the numerical results is given in the last section.

\section{Rosen-Morse potential}

\subsection{Energy-dependent Green function}
In this section we present another example --- a multilayered
heterostructure described by a confining potential $V(z)$ which
is chosen in the form of the Rosen-Morse potential
\begin{eqnarray}
V(z) &=& V_0\,{\rm tanh}^2\left({z\over L_{RM}}\right),\nonumber \\
V_0 &=& {\hbar^2 \over 2m L_{RM}^2}\ \kappa (\kappa+1).
\label{eq4.01}\end{eqnarray}
where $L_{RM}$ is the parameter close to the half-width of
the Rosen-Morse quantum well
and $\kappa$ is the dimensionless
parameter to govern the strength of the potential.

The summation (\ref{eq2.14}) over the quantum number $N$ can be
represented through the Green function which is known analytically
for the Rosen-Morse potential.
Namely, the second-order correction to the ground-state energy
can be written in the form
\begin{eqnarray}
\Delta_2 E &=& -\hbar\omega_{\mbox{\tiny LO}}\,\alpha\,{l_{RM}\over \sqrt{2}}
\int\limits_0^{\infty} dk_{\parallel} \int\limits_{-\infty}^{\infty} dz_a
\int\limits_{-\infty}^{\infty} dz_b\
e^{-k_{\parallel} |z_a-z_b|} \nonumber \\
&& \psi^*_1(z_a) \psi_1(z_b)\ G(z_a,z_b;E),
\label{eq4.02}\end{eqnarray}
where we made a scaling $z \to zL_{RM},\ \vec k \to \vec k/L_{RM}$ to use
dimensionless variables $z,\ \vec k$ and integrated
over $k_z$ and angles of $\vec k_{\parallel}$.
The dimensionless parameter
\begin{eqnarray}
l_{RM} = L_{RM} \sqrt{m\omega_{\mbox{\tiny LO}}\over \hbar}
\label{eq4.03}\end{eqnarray}
is the width of the confining potential in polaronic units
while the potential strength can now be written as follows:
\begin{eqnarray}
V_0 = \hbar\omega_{\mbox{\tiny LO}}{\kappa (\kappa +1)\over 2l_{RM}^2}.
\label{eq4.04}\end{eqnarray}

The quantity $G(z_a, z_b;E)$ is the Green function
of the dimensionless Hamiltonian (\ref{eq2.01})
which takes the form
\begin{eqnarray}
H''_{el,\perp} = -{1\over 2} {d^2\over dz^2} + {\kappa (\kappa +1)\over 2}
{\rm tanh}^2 z,
\label{eq4.05}\end{eqnarray}
that is $G(z_a, z_b;E) = \langle z_a|(H''_{el,\perp} -E)^{-1}|z_b\rangle$,
while $\psi_1(z)$ is the ground-state wave function of the potential
(\ref{eq4.05})
\begin{eqnarray}
\psi_1(z) = \left[{\Gamma(\kappa+1/2)\over \sqrt{\pi}\Gamma(\kappa)}\right]
^{1/2} {1\over {\rm cosh}^{\kappa} z}.
\label{eq4.06}\end{eqnarray}
The ground-state energy of the Hamiltonian (\ref{eq4.05}) is
given by
\begin{eqnarray}
E_1 = {\kappa\over 2}.
\label{eq4.07}\end{eqnarray}
The energy $E$ in Eq. (\ref{eq4.02}) reads as follows
\begin{eqnarray}
E = -{k^2_{\parallel}\over 2} - l_{RM}^2 + {\kappa \over 2}.
\label{eq4.08}\end{eqnarray}

The energy-dependent Green function of the system can be represented
in the form\cite{kleinert,flm}:
\begin{eqnarray}
&& G(z_a,z_b;E) =   \nonumber \\[3mm]
&& {\Gamma(\nu + \kappa+1) \Gamma(\nu-\kappa)\over
\Gamma^2(\nu+1)}\,
{1\over (4\,{\rm cosh}\,z_a\, {\rm cosh}\, z_b)^{\nu}}\times \nonumber \\
&&\phantom{q}_2F_1\left(\nu-\kappa, \nu+\kappa+1;\nu+1;{1-{\rm tanh}\, z_>
\over 2}\right) \times \nonumber \\
&&\phantom{q}_2F_1\left(\nu-\kappa, \nu+\kappa+1;\nu+1;{1+{\rm tanh}\, z_<
\over 2}\right),
\label{eq4.09}\end{eqnarray} \narrowtext\noindent
where $z_{>}\ (z_{<})$ denotes the maximum (minimum) of $z_a$ and $z_b$.
The parameter $\nu$ is defined by the relation
\begin{eqnarray}
\nu=\sqrt{-2\left(E-{\kappa (\kappa+1)\over 2}\right)}=
\sqrt{k^2_{\parallel} + \kappa^2 +2 l_{RM}^2}.
\label{eq4.10}\end{eqnarray}

The polaron effective mass can be represented in a similar way
\begin{eqnarray}
{\Delta_2 m\over m} &=&\alpha{l^3 \over 2\sqrt{2}}
\int\limits_0^{\infty} dk_{\parallel}\, k^2_{\parallel}
\int\limits_{-\infty}^{\infty} dz_a \int\limits_{-\infty}^{\infty} dz_b\
e^{-k_{\parallel}|z_a-z_b|} \nonumber \\
&& \psi_1^*(z_a) \psi_1(z_b)\ {\partial^2\over \partial E^2} G(z_a,z_b;E).
\label{eq4.11}\end{eqnarray}

To simplify numerical calculations we may replace the derivative
with respect to $E$ by the derivative with respect to $\nu$
\begin{eqnarray}
{\partial^2\over \partial E^2} = {1\over \nu^2}
{\partial^2\over \partial \nu^2} -{1\over \nu^3}
{\partial\over \partial \nu}
\label{eq4.12}\end{eqnarray}
and perform once the integration by parts. As the result, we arrive
at the following representation equivalent to Eq.~(\ref{eq4.11}):

\begin{eqnarray}
&& {\Delta_2 m\over m} =\alpha{l^3 \over 2\sqrt{2}}
\int\limits_0^{\infty} dk_{\parallel}\,
\int\limits_{-\infty}^{\infty} dz_a \int\limits_{-\infty}^{\infty} dz_b\
(1-k_{\parallel}|z_a-z_b|) \nonumber \\
&& e^{-k_{\parallel}|z_a-z_b|}
\psi_1^*(z_a) \psi_1(z_b)\
\left[-{1\over \nu} {\partial\over \partial \nu}\right] G(z_a,z_b;E).
\label{eq4.13}\end{eqnarray}

Note that $m, \alpha, \omega_{\mbox{\tiny LO}}$ in all these formulae
stand for the mean characteristics of the medium. The wave function
in their definitions is given by Eq.~(\ref{eq4.06}).
The numerical results are plotted
in Fig.~\ref{fig3} and discussed in the last Section.

\subsection{Effective width}

If we decide to compare the results for the rectangular and
Rosen-Morse potentials, we have to define a parameter which plays
the role of the effective width of the Rosen-Morse potential.
That is, this parameter (for which we use a notation $L$)
should be close to $2L_{RM}$ of Eq.~(\ref{eq4.01}) being also related
to the rectangular potential. We accept the following definition: let
us call the effective width of the Rosen-Morse
potential the width $L$ of the rectangular well of the same height
with the same ground-state energy in the
absence of the electron-phonon interaction (that is, at $\alpha =0$).
The advantage of this definition is that while calculating the polaron
binding energy for the Rosen-Morse and rectangular potentials
we subtract the same quantity in both the cases and can compare only
energy shifts due to the electron-phonon interaction.

The ground-state energy of a rectangular potential with the height $V_0$
and width $L$ is given by the relations
\begin{eqnarray}
E_{RC} &=& {\hbar^2 k^2 \over 2m},\nonumber \\
\mbox{\rm tan} \,{ {kL}\over 2} &=&
\sqrt{{V_0\over E_{RC}} - 1},
\label{eq4.15}\end{eqnarray}
while the RM ground-state energy looks like
\begin{eqnarray}
E_{RM} = {\hbar^2 \over m L_{RM}^2}\,{\kappa\over 2}
\label{eq4.16}\end{eqnarray}
and the height $V_0$ of the potential is given by Eq. (\ref{eq4.01}).
With the equality $E_{RM} = E_{RC}$ we arrive at the
relation between the parameter $L_{RM}$ of the Rosen-Morse potential and its
effective width defined as has been discussed:
\begin{eqnarray}
{L \over L_0} &=& 2\sqrt{\lambda}\,\mbox{\rm arctg}\sqrt{\lambda -1},
\nonumber \\
\lambda &=& \kappa +1 = {1\over 2}\left[1+
\sqrt{1+\left({2L_{RM}/L_0}\right)^2}\right].
\label{eq4.17}\end{eqnarray}
Here we introduce the distance scale
\begin{eqnarray}
L_0 = \sqrt{\hbar^2 \over 2m V_0}.
\label{eq4.18}\end{eqnarray}
The relation to the other dimensionless parameter $l_{RM}$
of Eq. (\ref{eq4.03}) is given by
\begin{eqnarray}
{L_{RM}\over L_0} = l\sqrt{2V_0 \over \hbar \omega_{\mbox{\tiny LO}} }.
\label{eq4.19}\end{eqnarray}
At small $L_{RM} \ll L_0$ we obtain $L \sim 2L_{RM}$ from Eq. (\ref{eq4.17}),
that is indeed the parameter $L_{RM}$ plays a role of the half-width
of the Rosen-Morse potential in this case. When $L_{RM} \gg L_0$,
it follows from Eq. (\ref{eq4.17}) that $L \sim \pi \sqrt{L_{RM}L_0}$.

The effective width $L$ defined in this subsection allows us to
apply the results for the rectangular potential to the Rosen-Morse
quantum well. The example is given in Fig.~\ref{fig3} where we
plotted also the energy and the mass shifts for the rectangular
potential vs. the parameter $L_{RM}$ related to $L$ as is described.

\section{Numerical results and discussion}

To proceed to the numerical calculations we need now the dependence
of medium parameters on the $AlAs$ mole fraction $x$.
At first we present the parametrization from the review by
Adachi\cite{adachi}:
\begin{mathletters}
\begin{eqnarray}
\alpha (z) = 0.068 + 0.058 x,
\label{eq5.01a}
\end{eqnarray}
\begin{eqnarray}
m(z) = m_e \cdot (0.0665 + 0.0835 x),
\label{eq5.01b}\end{eqnarray}
\begin{eqnarray}
\hbar\omega (z) =
(36.25 + 1.83 x + 17.12 x^2  -5.11 x^3)\ {\rm meV},
\label{eq5.01c}\end{eqnarray}
\label{eq5.01}
\end{mathletters}\noindent
which was used in numerical calculations by Hai, Peeters and
Devreese\cite{hpd2,hpd3}.
Here $m_e$ is the electron mass in vacuum and $m(z)$ --- its band mass
in the subsequent layer; the values of the electron-phonon coupling
constant $\alpha(z)$ and the LO-phonon frequency $\omega(z)$
are also related to this layer.

Some comments are to the point. The expression for the electron band mass
is nothing else but the linear
interpolation between the values $m=0.0665m_e$ for $GaAs$ and
$m=0.150m_e$ for $AlAs$. As to the LO-phonon frequency there are two
phonon modes with different frequencies $\omega^{(G)}(z)$
and $\omega^{(A)}(z)$ for the $GaAs$-like and $AlAs$-like modes
in $Al_x Ga_{1-x}As$ crystal. Experimental results of Ref.~\onlinecite{kim}
are interpolated by the following formulae:
\begin{mathletters}
\begin{eqnarray}
\hbar\omega^{(G)}(z) =
(36.25 - 6.55 x + 1.79 x^2)\ {\rm meV},
\label{eq5.02a}
\end{eqnarray}
\begin{eqnarray}
\hbar\omega^{(A)}(z) &=&
(44.63 + 8.78 x - 3.32 x^2)\ {\rm meV}.
\label{eq5.02b}
\end{eqnarray}
\label{eq5.02}
\end{mathletters}
Because the exact theory of the two-phonon interaction in alloys
where there are two-mode phonons present has not been reported,
Adachi suggested to use the effective phonon frequency
$\omega = (1-x)\omega^{(G)} + x \omega^{(A)}$, that is the linear
interpolation between these two modes.
Inserting here the expressions (\ref{eq5.02}) one arrives at the result
(\ref{eq5.01c}).

As to the interpolation formula (\ref{eq5.01a}) for the Fr\"ohlich
coupling constant $\alpha$, the situation seems to be a bit
inconsistent. Indeed, $\alpha$ depends on the values of the
static $\varepsilon_0$ and the high-frequency $\varepsilon_{\infty}$
dielectric constants:
\begin{eqnarray}
\alpha &=& {1 \over \hbar \omega}
{{\bar e}^2 \over \sqrt{2}} \sqrt{m \omega \over \hbar}
\left({1\over \varepsilon_{\infty}} - {1\over \varepsilon_{0}}\right)
\nonumber \\
&= & 116.643 \left({1\over \varepsilon_{\infty}}-
{1\over \varepsilon_{0}}\right) \sqrt{m\over m_e}
\sqrt{1\ {\rm meV}\over \hbar\omega}.
\label{eq5.03}\end{eqnarray}
Earlier measurements of $\varepsilon_0$ of $GaAs$ have yielded
widely different values ranging from 9.8 to 13.3
(see Ref.~\onlinecite{strzal} and references therein).
For instance, Kartheuser\cite{karth}
reports the values $\varepsilon_{\infty} =10.9$ and $\varepsilon_0=12.83$
and $\hbar \omega = 36.75\ {\rm meV}$ for $GaAs$.
This leads to the result $\alpha =0.068$, which is widely known and used
by many people.

On the other hand, Adachi used
the more recent results for $GaAs$\cite{samara}:
$\varepsilon_0=13.18\pm 0.40$ and $\varepsilon_{\infty}=10.89$,
and for $AlAs$\cite{fern}: $\varepsilon_0 = 10.06 \pm 0.04$
and $\varepsilon_{\infty}=8.16 \pm 0.02$.
This gives birth to his interpolation formulae\cite{adachi}:
\begin{mathletters}
\begin{eqnarray}
\varepsilon_0 = 13.18 - 3.12 x,
\label{eq5.04a}
\end{eqnarray}
\begin{eqnarray}
\varepsilon_{\infty} = 10.89 - 2.73 x,
\label{eq5.04b}
\end{eqnarray}
\label{eq5.04}
\end{mathletters}
Inserting formulae (\ref{eq5.01b}), (\ref{eq5.01c}) and (\ref{eq5.04})
into Eq.~(\ref{eq5.03}) Adachi declared the result $\alpha = 0.126$
for $AlAs$. Together with the value
$\alpha=0.068$ reported in Ref.~\onlinecite{karth} this leads
to the interpolation formulae (\ref{eq5.01a}). The problem is that
both these values for $\alpha$ do not follow from the parametrizations
mentioned above.

Taking the same values for $AlAs$ as Adachi did take
($m=0.150m_e,\ \hbar\omega = 50.09\ {\rm meV},\ \varepsilon_0 = 8.16,\
\varepsilon_{\infty} = 10.06$) we arrive at the result
$\alpha = 0.1477$. Moreover, if one takes the same interpolation
formulae (\ref{eq5.04}) at $x=0$ one obtains the value
$\alpha=0.0797$ for GaAs. That is, Adachi had to obtain
the formula
\begin{eqnarray}
\alpha(z) = 0.0797+0.0680 x
\label{eq5.05}\end{eqnarray}
as a linear interpolation between the values of $\alpha$
in $GaAs$ and $AlAs$. Note, that this formulae
can be presented in the form $\alpha(z) = 1.172(0.068+0.058x)$.
The expression between the brackets coincide (probably occasionally)
with the Adachi interpolation formulae for $\alpha$
(cf. Eq.~(\ref{eq5.01a})).
That is, the discrepancy of (\ref{eq5.01a})
and of our interpolation (\ref{eq5.05}) is about 17\%
and do not depend on $x$. To be consistent we have to
accept the parametrization (\ref{eq5.05}) in what follows.

For the confining potential we take the expression
derived from the band-gap energy fit of Ref.~\onlinecite{lee}
and used in Ref.~\onlinecite{hpd2,hpd3}:
\begin{eqnarray}
V(z) = 600 \cdot (1.155 x + 0.37 x^2 )\ {\rm meV}.
\label{eq5.06}\end{eqnarray}
Thus, we use the parametrization (\ref{eq5.01b}), (\ref{eq5.01c}),
(\ref{eq5.04a}),  (\ref{eq5.04b}), (\ref{eq5.05})
and the potential (\ref{eq5.06}) in our numerical calculations.

\begin{figure}[h]
   \hspace*{-2.6cm}
   \epsfysize=5.5in \epsfbox{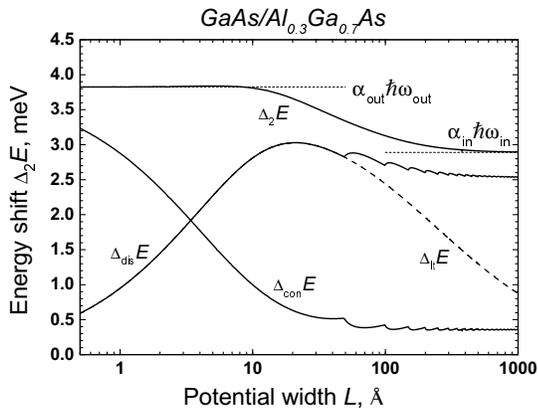}
   \vspace*{-7.5cm}
\caption{Total energy shift $\Delta_2 E$ is shown vs. the rectangular
potential width $L$ for $x=0.3$. The contribution of the discrete
($\Delta_{dis} E$) and continuous ($\Delta_{con} E$)
spectra are presented separately
as well as the result of the leading term approximation  ($\Delta_{lt} E$).}
\label{fig1}
\end{figure}

\begin{figure}[h]
   \hspace*{-0.0cm}
   \epsfysize=6in \epsfbox{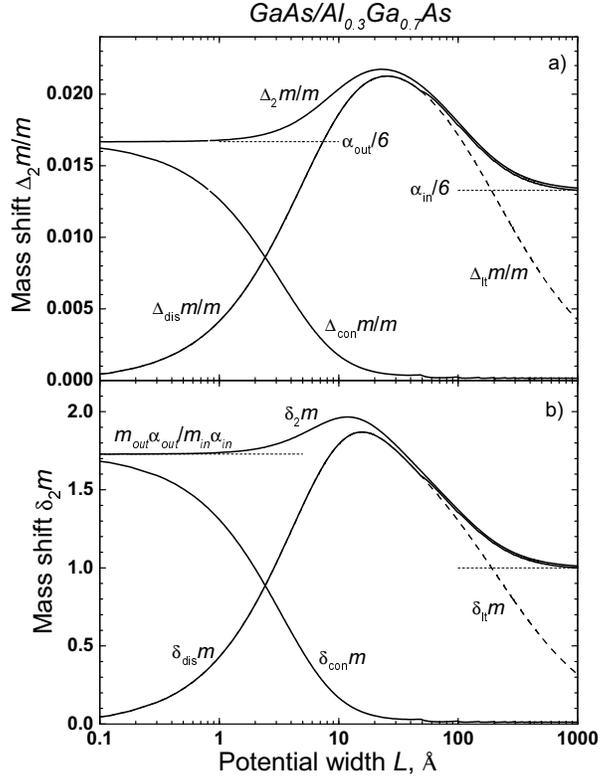}
   \vspace*{-4.0cm}
\caption{The relative shifts $\Delta_2m/m$
and $\delta_2 m = \Delta_2 m /\Delta_2 m_{in}$
of the effective polaron mass for the rectangular potential
at $x=0.3$. Contributions
of the discrete and continuous spectra are shown as well as the result
of the leading term approximation.}
\label{fig2}
\end{figure}

The results of our study for the rectangular
potential (which is formed by a layer of $GaAs/Al_x Ga_{1-x} As $)
are shown in Fig.~\ref{fig1} for the polaronic energy shift
and in Fig.~\ref{fig2} for the polaron effective mass
at the $AlAs$ mole fraction $x=0.3$.
The contribution of the discrete and continuous spectra are plotted
separately for this potential.  In Fig.~\ref{fig2}a
the relative mass shift
$\Delta_2m/m$ is shown where the mean mass $m$ also depends
on the potential width $L$. Thus, the ratio
$\delta_2m = \Delta_2m/\Delta_2 m_{in}$
of the mass shifts in the potential and in $GaAs$ is presented
also in Fig.~\ref{fig2}b for the same $AlAs$ mole fraction.
Evidently,
the asymptotics of this curve is equal to the unity at large $L$ and
to the ratio $m_{out}\alpha_{out}/m_{in}\alpha_{in}$ at $L \to 0$.

We may conclude that the continuous spectrum dominates at small
potential widths. At large widths its contribution could also be
significant although it is smaller than the contribution of the
discrete spectrum (especially in deep potential wells). We also confirm
the conclusion of the preceding papers that the leading term
approximation is not adequate to describe this system and leads to
wrong asymptotics at both small and large potential widths
(see the dashed lines in Figs~\ref{fig1}, \ref{fig2}).

\begin{figure}[h]
   \hspace*{-2.2cm}\vspace*{-2cm}
   \epsfysize=5.5in \epsfbox{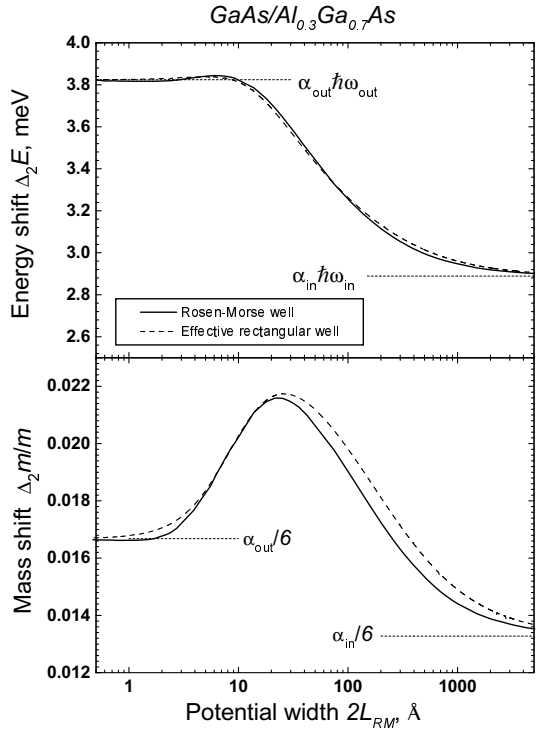}
   \vspace*{-1.5cm}
\caption{Polaron energy and effective mass shifts
for the Rosen-Morse potential (solid curves). The dashed lines present
results for the rectangular potential at $x=0.3$ as functions of
$2 L_{RM}$ recalculated from the width $L$ as is described in the text.}
\label{fig3}
\end{figure}

An example of a multilayered heterostructure is presented.
The results for the energy
and the effective mass for the polaron in the Rosen-Morse potential well
are shown in Fig.~\ref{fig3}. For the numerical
calculations we fix the value $V_0=227.9$ meV
in Eqs.~(\ref{eq4.01}), (\ref{eq4.04})
which corresponds to the limiting mole fraction
at large distances $x_{\infty}= \lim_{z\to \infty}x(z) = 0.3$.
Thus, we obtain the dependence of the mole fraction $x$
on the coordinate $z$:
\begin{eqnarray}
600 \cdot (1.155 x + 0.37 x^2 ) = 227.9\ {\rm tanh}^2 z.
\label{eq5.07}\end{eqnarray}
Now Eqs.~(\ref{eq5.01b}), (\ref{eq5.01c}), (\ref{eq5.05})
allow one to define the dependence of parameters
on the coordinate $z$ and to calculate the mean characteristics
of the heterostructure.

The calculations were completely different
in comparison with the rectangular potential: instead of the direct
summation over all intermediate
states we used the analytical expression for the Green function of the
Rosen-Morse potential. The results obtained demonstrate a similar behavior
which is also close numerically to the results for the rectangular
potential. The polaronic energy and mass shifts for the rectangular
quantum well are also plotted here (dashed line) vs. the Rosen-Morse
width $L_{RM}$ obtained from $L$ as is described above. We see that
both the energies almost coincide,
which gives the opportunity to approximate
different quantum wells by the rectangular potential. The discrepancy
in the effective mass is larger but not so crucial.
This serves also as an additional internal criterion of the
validity of our calculations.

Thus, we obtained a monotonous behavior of $\Delta_2 E$ between
the correct 3D limiting values $\alpha_{in}\hbar \omega_{in}$ and
$\alpha_{out}\hbar\omega_{out}$ both for the rectangular and
the Rosen-Morse potentials (see Fig.~\ref{fig1} and Fig.~\ref{fig3}a).
Actually the peaks are ``hidden" and they reveal themselves if we
plot the dimensionless energy shift
$\Delta_2 E/(\hbar\omega_{\mbox{\tiny LO}}\alpha)$ which has the same
3D limit (the unity) at both small and large potential widths.
But in the ``real" units (meV) the peaks are smoothed.

\begin{figure}
   \hspace*{-0.2cm}
   \epsfysize=4.5in \epsfbox{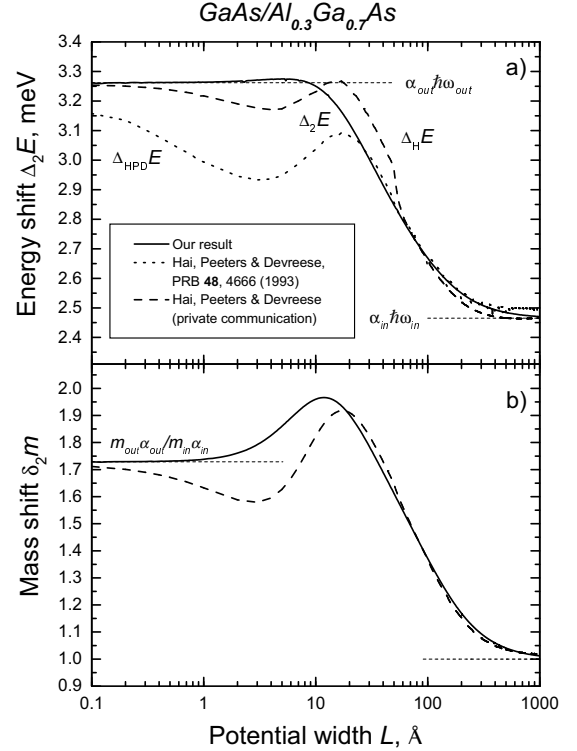}
\caption{Comparison of the results of the present paper (solid lines)
and those of Ref.~\protect\onlinecite{hpd2} (dotted line)
and of Ref.~\protect\onlinecite{hpd3} (dashed lines)
for the rectangular potential
generated by the $GaAs/Al_{0.3}Ga_{0.7}As$ heterostructure.
For this plot we used in our calculations the same parametrization
(\protect\ref{eq5.01}) and (\protect\ref{eq5.06})
as these authors did.}
\label{fig4}
\end{figure}

To compare our results with the calculations performed for the one-layer
heterostructure we refer to the papers\cite{hpd2,hpd3}
where the authors took into account
the contributions of different phonon modes as well as mass and dielectric
constant mismatches in the materials of the barrier and the well.
Note that the analytical formulas of Ref.~\onlinecite{hpd2}
contain a mistake ---
the wrong expression for the density of states. Namely,
in some parts of the continuous spectrum contribution the integration
is performed not over the wave vector $p$ but over the wave
vector $q$ (that is $\int_{V_0}^{\infty} dE_z/\sqrt{E_z}\, (\ldots)$
in the notations of that paper instead of the correct integration
$\int_{V_0}^{\infty} dE_z/\sqrt{E_z-V_0}\, (\ldots)$).
It is clear that
this mistake results in lowering of the resulting curve for the energy,
and the discrepancy
is larger when the energy is closer to the potential edge, that is at
small widths. This is just what we see in Fig.~\ref{fig4}a
comparing the result of Ref.~\onlinecite{hpd2}
(the curve $\Delta_{HPD} E$) with the new calculations
of the same authors (the curve $\Delta_{HPD} E$)
which came to our knowledge when the present paper
was already submitted for the publication.

Thus, our model does not reproduce the
more complicated structure with the peak and the dip
which was obtained in Ref.~\onlinecite{hpd3}. Some hints on the
existence of  peaks
can also be seen in our plots but the maximal values are
so close to the asymptotics that the peaks are almost invisible.
Probably, the dip appears because of
the presence of several phonon modes (bulk, interface, etc.).
At widths $L \geq 50\ \AA$ our results for the energy
practically coincide with those of Ref.~\onlinecite{hpd3}.
The discrepancy at smaller widths seems to be more crucial.
But the difference
between the values in the peak and in the dip for the curve
$\Delta_H E$ in Fig.~\ref{fig4}a
is about 0.1 eV (3\%). This phenomena hardly can
be seen experimentally and this discrepancy
is in the limits of the accuracy of our model estimated above.
This gives indeed a strong support to our model and we
may conclude that the latter provides us with the rather accurate
approximation and can be used for more complicated calculations
in multilayered heterostructures.

As to the shift of the electron band mass we found clear peaks
for both the rectangular and the Rosen-Morse potentials
(see Figs.~\ref{fig2}, \ref{fig3}). As is seen in
Fig.~\ref{fig2}
the effective mass shift for the polaron in the rectangular quantum well
has a peak at $L \approx 20 \AA\ (x=0.3)$.
Calculations show also that the larger is $x$ the smaller is the
potential width corresponding to the peak.
For the Rosen-Morse potential at $x_{max}=0.3$ the peaks in the
effective mass occur at $2L_{RM} \approx 20 \AA$.
Note, that again the authors
of Ref.~\onlinecite{hpd3} obtained curves with peaks and dips
in contrast with our results (see Fig.~\ref{fig4}b).
The maximal discrepancy for the mass is about 11\% at $L \approx 3 \AA$
which is beyond the region available for experiments.
Our results are very close to those of Ref.~\onlinecite{hpd3}
at $L \geq 10 \AA$ and practically coincde with them at $L \geq 20 \AA$.

\begin{figure}
   \hspace*{-0.2cm}
   \epsfysize=6in \epsfbox{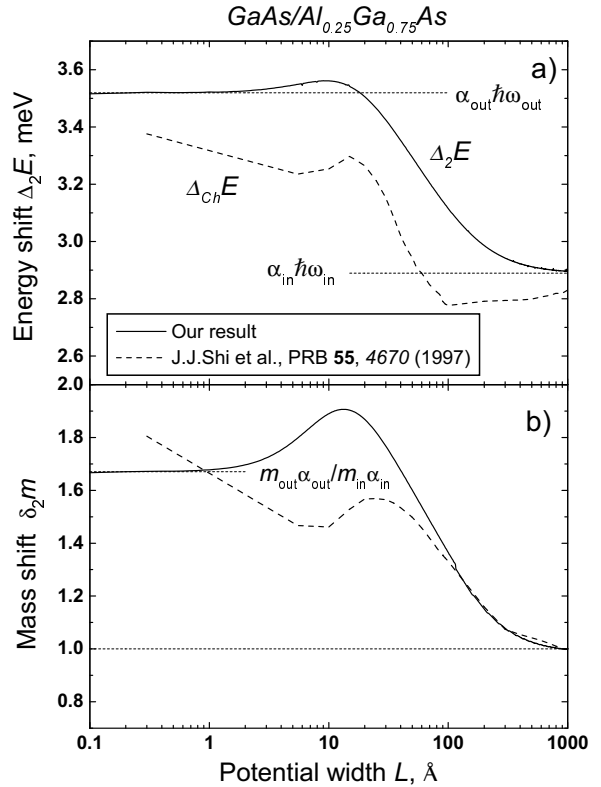}
   \vspace*{-4.0cm}
\caption{Comparison of the results of the present paper and
those of Ref.~\protect\onlinecite{china} for the rectangular potential
generated by the $GaAs/Al_{0.25}Ga_{0.75}As$ heterostructure.
For this plot we used the parametrization of these authors as
is described in the text.}
\label{fig5}
\end{figure}

To compare our results with those of Ref.~\onlinecite{china}
we need now another parametrization used by these authors
(although they refer also to the paper by Adachi\cite{adachi}).
Namely, they took a slightly different expression
for the confining potential:
\begin{eqnarray}
V(z) = 600 \cdot (1.266 x + 0.26 x^2 )\ {\rm meV},
\label{eq5.08}\end{eqnarray}
which follows from the band gap of Ref.~\onlinecite{abram}.
Furthermore, instead of the effective LO-phonon frequency
they used the expression (\ref{eq5.02a}) for the energy of
the $GaAs$-like phonons. The Fr\"ohlich coupling constant $\alpha$
was calculated then using also the parametrization (\ref{eq5.01b})
and (\ref{eq5.04}).
Note, that these numerical calculation, as we found,
can be approximated by the interpolation formula
\begin{eqnarray}
\alpha(z) =0.0797 + 0.0772 x + 0.0295 x^2.
\label{eq5.09}\end{eqnarray}
The results of the comparison are shown in Fig.~\ref{fig5}
(we used in our calculations for this plot the same
parametrization as was used in Ref.~\onlinecite{china}).

The curve $\Delta_{Ch}E$ in Fig.~\ref{fig5}a
for the energy shift taken from Ref.~\onlinecite{china}
has also a small dip (qualitatively similar to this of
Ref.~\onlinecite{hpd3}). But the discrepancy between energy shifts
is much more drastic in this case, and
we have no explanation for this.

It is clear that
at large potential width only a bulk phonon mode inside the
quantum well contributes so these curves should have
the same limiting value $\alpha_{in}\omega_{in}$. Numerically
we found $\alpha_{in} = 0.0797$ and $\omega_{in} = 36.25$ meV,
so $\alpha_{in}\omega_{in} = 2.89$ meV. Moreover, the behavior
of the curves at large $L$ should be qualitatively and quantitatively
the same which was the case when we compared our model
with Refs.~\onlinecite{hpd2,hpd3}.
In contrast with our model and the cited results by Hai, Peeters
and Devreese
the curve $\Delta_{Ch} E$ in Fig.~\ref{fig5}b approaches the asymptotics
from below and the subsequent mechanism remains unclear.
On the other hand there are some reasons why the curve have to
approach its asymptotics from above. Indeed, at large potential width
the particle does not feel yet the finite height of the potential,
and the energy shift takes the same value as in the infinitely high
potential which is a bit larger than the free polaron energy.

As to the opposite limit of the small width of the confining potential,
it is surprising that the asymptotic value is not reached even
at $L \sim 0.3\ \AA,$ as is found in Ref.~\onlinecite{china}.
Numerically we obtained $\alpha_{out} = 0.1014$ in this scheme
and $\omega_{out}= 34.72$ meV, so  $\alpha_{out}\omega_{out} = 3.52$ meV.

Both asymptotic values coincide with what was obtained by the authors
of Ref.~\onlinecite{china}.
Looking at the behavior of the mass shift, we see that both curves
coincide at large widths as it should be. At widths smaller than 100 \AA\
the discrepancy becomes evident. But we may conclude that
something is wrong with the numerical job
of Ref.~\onlinecite{china} because their curve approaches the wrong
limit at $L \to 0$. Indeed, in this limit
the asymptotical value of the plotted ratio should
be equal to $\alpha_{out}m_{out}/ \alpha_{in}m_{in}$.
As it follows from our analysis of the energy shift,
we obtained the same values for the Fr\"ohlich coupling constants.
The values for the band masses follow from Eq.~(\ref{eq5.01b}):
$m_{in} = 0.0665m_e$ and $m_{out} = 0.0874m_e$ at $x=0.25$.
Then, the asymptotical value of the plotted ratio should be equal
to $1.67$, instead of 1.83 what was found in Ref.~\onlinecite{china}
That is, the discrepancy is about 10\% in this limit and we cannot explain
its origin as well.

It would be highly desirable to include a comparison of our
results and the results by other authors
with corresponding experiments. To the best of our
knowledge, no such experiments do exist at the moment.

\section{Conclusions}

To conclude, we suggested an approximate
model to describe a multilayered
$GaAs/Al_x Ga_{1-x} As$ heterostructure as an effective medium
with one (bulk) phonon mode.
The fundamental entity is the confining
potential generated by these layers which we take into account
explicitly. Then we calculate the mean characteristics of the electron
in the effective medium (such as its band mass, phonon frequencies etc.)
which depend on the form of the confining potential.
With these parameters
we calculated the energy and the effective mass of a polaron
confined to a quasi-2D quantum well $GaAs/Al_x Ga_{1-x} As$
for different $AlAs$ mole fractions. The calculations include
the full energy spectrum as intermediate states.
Peaks are found for the effective mass at some potential widths while
the energy demonstrates rather monotonous behavior between
the correct 3D-limits. Finally, some discrepancies
in the interpolation formulae for the experimental results are discussed
as well as discrepancies with the previously obtained
theoretical results. We demonstrated that our model gives practically
the same (or very close) results
as the explicit calculations of Ref.~\onlinecite{hpd3}
for potential widths $L \geq 10 \AA$.

\acknowledgments
We thank J.~T. Devreese, V.\ N. Gladilin, G.\ Q. Hai, H. Leschke, V.~M. Fomin,
F.~M.~Peeters, E.~P. Pokatilov, and J.~W\"usthoff
for useful discussions and valuable remarks and advices.
Special thanks
are to the authors of the paper \cite{hpd3}
for making their results available to us prior to publication
and to the authors of the paper \cite{china}
who kindly provided us with their data-files.

M.O.D. and M.A.S. are grateful to Dortmund University
for the kind hospitality during their visits to Germany.
Financial support of the Heisenberg-Landau
program (Germany-JINR collaboration in theoretical physics)
and Deutsche Forschungsgemeinschaft  (Graduiertenkolleg GKP 50/2)
is gratefully acknowledged.

\end{document}